%% file: varLUTGRAND.tex
\documentclass[journal,10pt]{IEEEtran}

\usepackage{cite}
\usepackage{epsfig}
\usepackage{epstopdf}
\usepackage{graphicx}
\usepackage[dvipsnames]{xcolor}
\usepackage{tikz}
\usetikzlibrary{spy,backgrounds,arrows,positioning,shapes.geometric,circuits.logic.US}
\usepackage{pgfplots}
\usepackage{multirow}
\usepackage{upgreek}
\usepackage{amssymb}
\usepackage{amsmath}
\usepackage{cases}
\usepackage{url}
\usepackage{pifont}
\usepackage{ctable}
\usepackage{bm}
\usepackage{amsthm}

\usepackage{tabularx}
\usepackage{booktabs}
\usepackage{filecontents}
\usepackage{subcaption}
\newcolumntype{Y}{>{\centering\arraybackslash}X}

\theoremstyle{definition} 
\theoremstyle{definition} 
\theoremstyle{definition} 
\theoremstyle{definition}

\usepackage[linesnumbered, ruled, vlined, noend]{algorithm2e}

\usepackage{algorithmic}

\usepackage{array}

\newcommand{\fixme}[2]{\ifx&#2&{\leavevmode\color{red}#1}\else{\leavevmode\color{red}FIXME\{}#1{\leavevmode\color{red}\}}\footnote{{\leavevmode\color{red}#2}}\PackageWarning{Fixme}{#1: #2}\fi}

\newcommand{\newstuff}[2]{\ifx&#2&{\leavevmode\color{blue}#1}\else{\leavevmode\color{blue}FIXME\{}#1{\leavevmode\color{blue}\}}\footnote{{\leavevmode\color{blue}#2}}\PackageWarning{Newstuff}{#1: #2}\fi}

\title{A 95.5Gb/s 29.6ns worst-case latency \\ORBGRAND decoder for 6G xURLLC}

\author{\IEEEauthorblockN{Carlo~Condo}} 

\begin{document}

\maketitle
\begin{abstract}
Ultra-Reliable Low-Latency Communications (URLLC) in both 5G and 6G demand high throughput and short latency with low error rates. 
Guessing Random Additive Noise Decoding (GRAND) and Ordered Reliability Bits GRAND (ORBGRAND) are powerful universal decoding algorithms that work well with short, high-rate codes. 
As short forward error correcting codes can help limiting latency, and code unification in 6G calls for flexible, possibly code-agnostic decoders, GRAND and ORBGRAND are well suited to tackle 6G URLLC.
This work proposes a ultra-high, constant speed ORBGRAND decoder architecture with very low worst-case and average latency.
Compared to a baseline architecture, through out-of-order output, aggressive clock gating, and selective programmability, the decoder reduces area, power, and average latency by 15.5\%, 19.4\%, and 56\%, respectively.
In 3nm FinFET technology, it achieves a constant throughput of 95.49Gb/s, with 29.59ns worst-case latency and 13.02ns on average.

\end{abstract}


\IEEEpeerreviewmaketitle

\section{Introduction} \label{sec:intro}

The Ultra-Reliable Low-Latency Communication (URLLC) scenario in 5G describes fast, high-performance, low-latency links for a variety of mission-critical applications, including remote surgery, smart grid monitoring, and vehicle-to-vehicle communications.
The upcoming 6G standard will push the reliability and latency constraints of URLLC-like communications even further in its next-generation URLLC (xURLLC), along with covering some key performance indicators required by sensitive applications, where 5G URLLC falls short \cite{xURLLC}.

Forward error correction (FEC) is a key component of the physical layer, requiring complex operations and often contributing high power consumption and long latency.
As such, short codes are desirable for URLLC and xURLLC in order to mitigate the decoder latency \cite{URLLCshort}, and the search for low-power, low-latency decoding algorithms and decoder architectures is of interest for industry and academia alike.
Along with these requirements, the design of FEC in 6G is foreseen to strive towards code unification \cite{6Gcodes}, where a small set of code families or even a single code type can cater to a wide range of communication requirements.

As FEC decoders are usually far more complex than encoders, an important step towards code unification can be achieved via code-agnostic decoding. 
A single decoder with the ability to decode not only different code lengths and rates, but even code types, is an efficient way to limit power consumption and silicon footprint.
Guessing Random Additive Noise Decoding (GRAND) \cite{GRAND_first} is a universal decoder can achieve maximum-likelihood (ML) or near-ML decoding with limited complexity, and that can be used to decode any type of code. 
It has been shown to work well with short, high-rate codes.
Ordered Reliability Bits GRAND (ORBGRAND) \cite{ORBGRAND_first} substantially improves the error-correction performance of GRAND while remaining implementation-friendly.
Thanks to its attractiveness for practical applications, enhanced versions of ORBGRAND can be found in literature \cite{iLWO,LGRAND}, along with hardware implementations \cite{LUTGRAND,ORBCHIP,ORB_gross}.

Within this framework, this work proposes an ultra-high speed ORBGRAND decoder architecture with very low worst-case and average latency. 
With the architecture presented in \cite{LUTGRAND} as a baseline, the new decoder leverages out-of-order output, selective programmability, and aggressive clock gating to improve area occupation, power consumption, and average latency, without sacrificing any throughput or error-correction performance at realistic working points.

\section{Preliminaries} \label{sec:prel}


Given a binary linear block code code $\mathcal{C}$, let $\mathbf{G}$ bet the $k \times n$ generator matrix and $\mathbf{H}$ the $(n-k) \times n$ parity check matrix.
An $n$-bit codeword $\mathbf{x}$ is then obtained as $\mathbf{x}=\mathbf{u}\cdot \mathbf{G}$, where $\mathbf{u}$ is a $k$-bit information vector.
There are $2^k$ possible $\mathbf{x}$ in the codebook of $\mathcal{C}$, satisfying the following:
\begin{equation}
\forall \mathbf{x} \in \mathcal{C},\mathbf{H} \cdot \mathbf{x}^{\rm T} = \mathbf{0}~,
\end{equation}
where $\mathbf{0}$ is the all-zero vector. 
Let us transmit $\mathbf{x}$ through a noisy channel, and let us call $\mathbf{y}$ the received logarithmic likelihood ratio (LLR) vector. Then ${\rm HD}(\mathbf{y}) = \mathbf{x} \oplus \mathbf{e}$ is the hard decision of $\mathbf{y}$, with $\mathbf{e}$ being the error pattern applied by the channel.
Errors are detected whenever $\mathbf{H} \cdot {\rm HD}(\mathbf{y})^{\rm T} \neq \mathbf{0}$.

GRAND \cite{GRAND_first} attempts the retrieval of the error pattern applied by the channel by querying the codebook through $\mathbf{H} \cdot ({\rm HD}(\mathbf{y}) \oplus \mathbf{e})^{\rm T}$ for different test patterns $\mathbf{e}$. 
In case the result is $\mathbf{0}$, decoding is successful and vector $\hat{\mathbf{y}}={\rm HD}(\mathbf{y}) \oplus \mathbf{e}$ is returned, otherwise a new query is performed with a different $\mathbf{e}$.
To limit complexity at a small performance cost GRAND with abandonment  \cite{GRAND_first} stops after $Q_{max}$ codebook queries. 

ML decoding is achieved when scheduling the error patterns in decreasing order of probability. 
For binary symmetric channels it is sufficient to try $\mathbf{e}$ with increasing Hamming weight $HW$, but this schedule suffers strong performance degradation with additive white Gaussian noise channels (AWGN). 
The ORBGRAND algorithm \cite{ORBGRAND_first} can infer a refined error-pattern schedule by using the soft information vector $\mathbf{y}$ instead of ${\rm HD}(\mathbf{y})$ only.
Vector $\mathbf{y}$ is sorted in ascending order of reliability (i.e. increasing magnitude for LLRs), resulting in the index permutation $\pi$ and the sorted vector $\pi(\mathbf{y})$. 
In \cite{ORBGRAND_first}, the error patterns $\mathbf{e}$ applied to $\pi(\mathbf{y})$ are scheduled in ascending \emph{logistic weight order} (LWO). 
Given the ordered vector $\mathbf{v} = (v_0,\dots,v_{HW-1})$ containing the indices of the nonzero entries of $\mathbf{e}$, and its length $HW$, then the logistic weight $LW$ of $\mathbf{e}$ can be computed as
\begin{equation}\label{eq:LW}
LW(\mathbf{e}) = \sum_{i=0}^{HW-1} (v_i+1)~.
\end{equation}

The work in \cite{iLWO} proposes the improved logistic weight order (iLWO) schedule, that yields better error-correction performance in low noise conditions.
The weight of each error pattern $\mathbf{e}$ is computed as: 
\begin{equation}\label{eq:iLW}
iLW(\mathbf{e}) = \sum_{i=0}^{HW-1} (i+1)\cdot(v_i+1)~.
\end{equation}
Unlike LWO, iLWO favors patterns with low $HW$ and potentially high $LW$, following the observation that at low enough block error rate (BLER), high-$HW$ patterns are less likely to decode successfully.
The look-up-table (LUT) aided scheduling, in which the first $Q_{LUT}$ error patterns scheduled are observed empirically and stored in a LUT, is used with iLWO in \cite{LUTGRAND}, further improving performance.

GRAND-based decoding is code agnostic. 
Without loss of generality, the examples in the rest of the manuscript consider a two-error-correcting Bose-Chaudhuri-Hocquenghem (BCH) code with $n=127$, $k=113$, labeled BCH(127,113,2). 
The BCH generator polynomial defined on GF($2^7$) is $g_\mathcal{C}=$ 0x7761, and the field generator polynomial is $g_\mathcal{F}=$ 0x91. 

\begin{figure*}[t!]
	\centering
	\includegraphics[width=\linewidth]{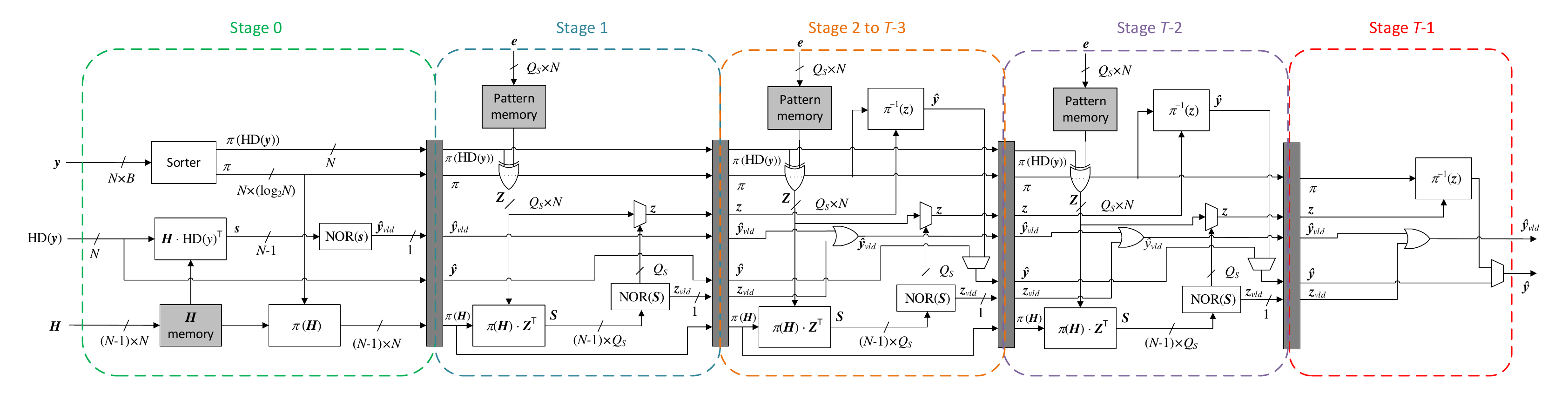}
	\caption{ORBGRAND decoder architecture in \cite{LUTGRAND}.}
	\label{fig:DEC}
\end{figure*}

In \cite{LUTGRAND}, a fixed-latency ORBGRAND decoder architecture was proposed. 
It is based on a pipelined architecture with $T$ stages, where a total of $Q_{max}$ error patterns are attempted over $T$-2 stages.
The decoder is shown in Figure \ref{fig:DEC}. 
Stage 0 contains a sorter used to obtain $\pi$, $\pi(\mathbf{y})$, and the permuted parity check matrix $\pi(\mathbf{H})$. 
Moreover, the syndrome of the input vector is computed to see if $\mathbf{y}$ is a valid codeword.
At stages $1\le t < T$ a LUT stores $Q_S$ test error patterns, that are applied to $\pi(\mathbf{y})$ in parallel: the resulting vectors are multiplied to $\pi(\mathbf{H})$ see if any belongs to $\mathcal{C}$. 
In case of success, the corrected vector $\hat{\mathbf{y}}$ is retrieved in the next stage, and the subsequent ones are clock gated to save power. 
If no valid codeword is found, the decoding continues.
This decoder architecture is the starting point of the remainder of the paper.

\section{Proposed decoder architecture}\label{sec:newarch}

\begin{figure}[t!]
	\centering
	\includegraphics[width=\linewidth]{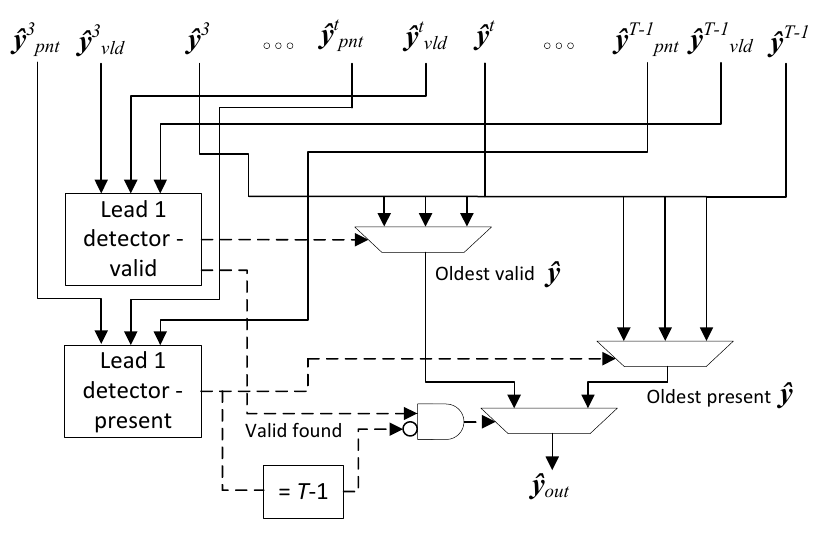}
	\caption{Output selection circuit.}
	\label{fig:outsel}
\end{figure}

In wireless communication systems, information is divided into packets before being encoded. 
This means that packet headers are contained within the encoded bits, to enable the receiver to reconstruct the transmitted information after decoding also in case of out-of-order reception.
This general feature enables a particular improvement to the architecture proposed in \cite{LUTGRAND}, and to unrolled decoders in general \cite{UNROL2}.

The decoder in \cite{LUTGRAND} yields high fixed throughput and low latency. 
With its unrolled architecture, the low average number of codebook queries required for decoding is leveraged not to have low average latency, but to clock gate the later decoding stages and reduce power consumption.
The decoder can yield similar power consumption and the same worst case latency by allowing out-of-order output, thus also benefiting from a low average latency, while at the same time guaranteeing fixed throughput.
Instead of the output being taken from stage $T-1$, it can be taken from any stage that contains, ideally, a valid codeword. 
If no valid codewords are present in the pipeline, an invalid one needs to be output anyway for constant throughput.

\begin{algorithm}[t!]
\newcommand\mycommfont[1]{\footnotesize\ttfamily\textcolor{blue}{#1}}
\SetCommentSty{mycommfont}
\def\HiLi{\leavevmode\rlap{\hbox to \hsize{\color{yellow!50}\leaders\hrule height .8\baselineskip depth .5ex\hfill}}}
\def\HiLiO{\leavevmode\rlap{\hbox to \hsize{\color{orange!50}\leaders\hrule height .8\baselineskip depth .5ex\hfill}}}
\def\HiLa{\leavevmode\rlap{\hbox to \hsize{\color{blue!30}\leaders\hrule height .8\baselineskip depth .5ex\hfill}}}
		\SetKwInOut{Input}{input}%
		\SetKwInOut{Output}{output}%
		\SetKw{Return}{return}
		\SetAlgoLined
		\Input{$\mathbf{\hat{y}}_{pnt}$, $\mathbf{\hat{y}}_{vld}$, $\mathbf{\hat{y}}$, $T_{fill}$}
		\Output{$\mathbf{\hat{y}}_{out}$, updated $\mathbf{\hat{y}}_{pnt}$, updated $\mathbf{\hat{y}}_{vld}$}
		Fill decoder pipeline up to $T_{fill}$;\\
		valid$\_$found $\leftarrow 0$;\\
		\If{$\mathbf{\hat{y}}^{T-1}_{pnt} = 1$}{
	    $\mathbf{\hat{y}}_{out} \leftarrow \mathbf{\hat{y}}^{T-1}$;\\
		}
		\Else{
		\For{$t =T$-$1\dots 3$}{
		\If{$\mathbf{\hat{y}}^t_{vld} = 1$}{
		$i_{vld} \leftarrow t$\tcp*{Oldest valid codeword}
		valid$\_$found $\leftarrow 1$;\\
		{\bf break};\\
		}
		}
		\For{$t =T$-$1\dots 3$}{
		\If{$\mathbf{\hat{y}}^t_{pnt} = 1$}{
		$i_{pnt} \leftarrow t$\tcp*{Oldest vector}
		{\bf break};\\
		}
		}
		\If{{\rm valid$\_$found} $= 1$}{
	    $\mathbf{\hat{y}}_{out} \leftarrow \mathbf{\hat{y}}^{i_{vld}}$;\\
		$\mathbf{\hat{y}}^{i_{vld}}_{vld},~\mathbf{\hat{y}}^{i_{vld}}_{pnt} \leftarrow 0$;\\
		}
		\Else{
	    $\mathbf{\hat{y}}_{out} \leftarrow \mathbf{\hat{y}}^{i_{pnt}}$;\\
		$\mathbf{\hat{y}}^{i_{pnt}}_{vld},~\mathbf{\hat{y}}^{i_{pnt}}_{pnt} \leftarrow 0$;\\
		}
		}
		\Return{$\mathbf{\hat{y}}_{out}$, $\mathbf{\hat{y}}_{vld}$, $\mathbf{\hat{y}}_{pnt}$}
	\caption{Proposed output selection criterion.}\label{alg:outsel}
\end{algorithm}

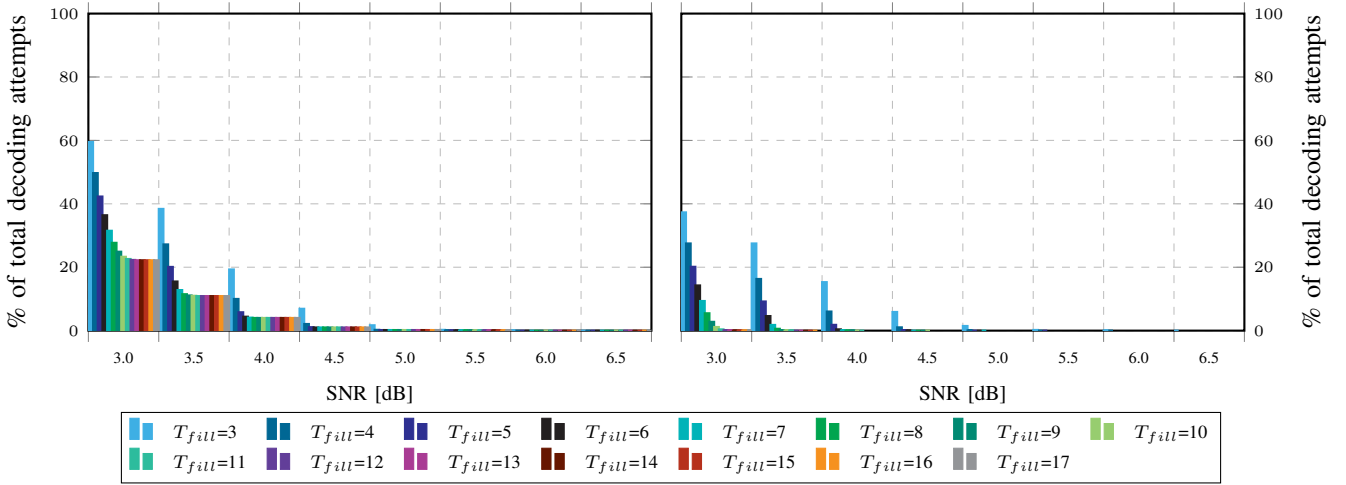
\begin{figure*}[t!]
    \centering
    \begin{minipage}{.5\textwidth}
        \centering
		  \input{figures/TotalErrDistr.tikz}
		        \end{minipage}%
    \begin{minipage}{0.5\textwidth}
        \centering
		    \input{figures/AddErrDistr.tikz}
   \end{minipage}
    \ref{TotErr}
    \caption{Total and additional failed decoding w.r.t. baseline ($T_{fill}=T-1=17$), for BCH(127,113,2) and different $T_{fill}$.}
    \label{fig:err}
\end{figure*}

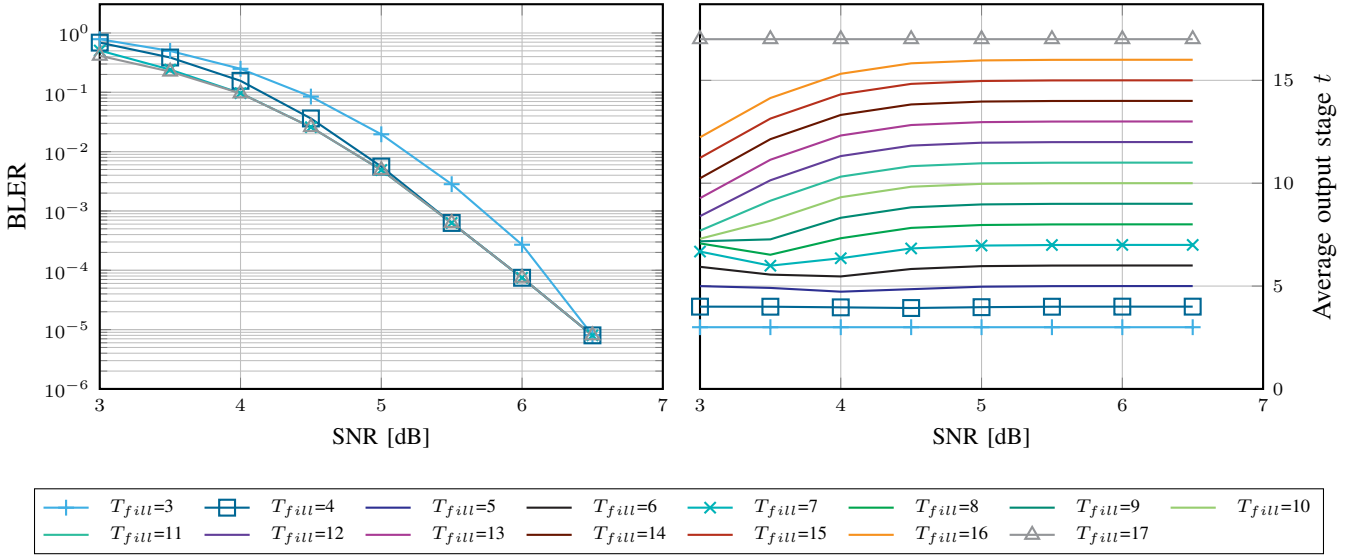
\begin{figure*}[t!]
    \centering
    \begin{minipage}{.5\textwidth}
        \centering
        		  \vspace{12pt}
		  \input{figures/BLER_Tfill.tikz}
		        \end{minipage}%
    \begin{minipage}{0.5\textwidth}
        \centering
		    \input{figures/Complexity_Tfill.tikz}
   \end{minipage}
    \ref{Comp_tfill}
    \caption{Effect of different $T_{fill}$ on the BLER and average output $t$ of BCH(127,113,2).}
    \label{fig:tfill}
\end{figure*}

Algorithm \ref{alg:outsel} describes the idea.
At each stage $t$ signal $\mathbf{\hat{y}}^t_{vld}$ indicates if vector $\mathbf{\hat{y}}^t$ is a valid codeword: in case it is not, $\mathbf{\hat{y}}^t$ is equal to ${\rm HD}(\mathbf{y})$.
Signal $\mathbf{\hat{y}}^t_{pnt}$ indicates instead if $\mathbf{\hat{y}}^t$ is present or it has been popped out already. 
The decoder pipeline is initially filled up to stage $T_{fill}$.
Then, at every clock one of the $\mathbf{\hat{y}}^t$ with $3\le t < T$ is output. 
In case a vector has reached the end of the pipeline, it is chosen as the output regardless of its status (Line 3-4).
The highest stage $t$ that contains a valid codeword is identified in Line 6-10. 
At the same time, the highest stage $t$ that contains a vector is identified in Line 11-14. 
In case a valid codeword was found in Line 6-10, it is selected as the output (Line 15-17). 
In case no valid codewords are present in the pipeline, the oldest vector is sacrificed and output (Line 18-20). 
Line 6 and 11 stop the search for viable candidate outputs at $t=3$, thus excluding $t=\{0,1,2\}$: while all of the $T$ stages can at any time contain a correct codeword, in the case of $t=0,1$ this is the input vector $\mathbf{y}$, and in $t=2$ $\mathbf{\hat{y}}$ is available only at the end of a long critical path.
For $T_{fill}=T-1$ the decoder is exactly the one presented in \cite{LUTGRAND}.
Outputting the present vector at the highest $t$ in case no valid codewords are found minimizes the chance of future decoding failures, since it has the smallest chance of being corrected.

The architecture of the output selection circuit is detailed in Figure \ref{fig:outsel}. 
A lead-1 detector identifies which is the oldest valid $\mathbf{\hat{y}}$ in the decoder pipeline by observing $\mathbf{\hat{y}}^t_{vld}$ with priority to higher indices. 
This module also returns a ``valid found" flag, that is set to 0 if $\mathbf{\hat{y}}^t_{vld}=0~\forall~3\le t<T$.
In parallel, another lead-1 detector identifies the oldest present $\mathbf{\hat{y}}$ by observing $\mathbf{\hat{y}}^t_{pnt}$, again with priority to higher indices.
If there are no valid codewords, or if $\mathbf{\hat{y}}^{T-1}_{pnt}=1$, then the oldest present $\mathbf{\hat{y}}$ is chosen as the output, otherwise the oldest valid $\mathbf{\hat{y}}$ is. 
Depending on the chosen output, the $\mathbf{\hat{y}}^t_{pnt}$ and possibly $\mathbf{\hat{y}}^t_{vld}$ are updated and passed to the following pipeline stage.

The value of $T_{fill}$ is programmable, and allows to tune the desired trade-off between average latency and error-correction performance to different codes and channel conditions. 
Figure \ref{fig:err} shows how different $T_{fill}$ affect the percentage of total failed decoding attempts (left) and the percentage of additional failed decoding attempts with respect to the baseline in \cite{LUTGRAND} (right), for BCH(127,113,2).
While $T_{fill}<8$ can have a noticeable impact on the number of failed decoding attempts at low SNR, this metric goes to zero as the channel conditions improve for almost all $T_{fill}$. 
BLER and average latency, measured as the average stage $t$ at which codewords are output, are shown in Figure \ref{fig:tfill}.
The left graph indicates that $T_{fill}=3$ degrades the decoder performance by up to 0.35dB, nevertheless converging to the $T_{fill}=T-1=17$ baseline at BLER$<10^{-5}$. 
$T_{fill}=4$ matches the baseline performance at BLER$<10^{-3}$, and the BLER of $T_{fill}=7$ is indistinguishable from that of $T_{fill}=17$ even at low SNR.
The right graph shows that the average latency of each $T_{fill}$ quickly converges to $T_{fill}$ itself as the SNR increases.
Consequently, for this case study $T_{fill}=4$ guarantees the lowest average latency with no error-correction performance degradation at any realistic BLER.
Extensive comparisons between LA-iLWO and other state of the art ORBGRAND schedules can be found in \cite{LUTGRAND}.

\subsection{Hardwired patterns}

In \cite{LUTGRAND} the LUT-aided error pattern schedule, that combines $Q_{LUT}$ error patterns observed empirically with any other schedule, was used with iLWO (LA-iLWO) for its superior performance in low noise conditions.
The $Q_{max}-Q_{LUT}$ error patterns generated with iLWO exclude the intersection with the observed $Q_{LUT}$ patterns.
For the case study in \cite{LUTGRAND}, where $n=127$, $Q_{LUT}=512$, and $Q_{max}=2^{13}$, the intersection accounts for 504 patterns.
This means that $Q_{LUT}$ contributes $512-504=8$ test patterns that would not be attempted within $Q_{max}$ queries in iLWO.
These few patterns are responsible for the improved error correction performance of LA-iLWO with respect to iLWO.
The baseline decoder allows all $Q_{max}$ error patterns to be programmable, but this comes at the substantial complexity cost of $Qs\times N$ registers per stage.
To reduce the complexity and power consumption of the decoder, in the proposed architecture only 32 non-iLWO patterns remain programmable, while the other $Q_{max}-32$ are implemented as hardwired LUTs. 
This choice puts an upper bound to the decoder BLER that equals that of iLWO with $Q_{max}=2^{13}-32$, while at the same time allowing 32 high-probability error patterns to be tailored to the code. 
Whereas 8 programmable patterns would be enough to have the same performance as \cite{LUTGRAND}, 32 grant a higher degree of flexibility while still allowing for substantial complexity reduction.

Another feature of the previous version of the decoder is that any code length $n$ lower or equal than a maximum $N$ can be decoded, with the programmable error patterns guaranteeing that $Q_{max}$ valid patterns are always present.
With the proposed hardwired patterns, any one attempting to flip a bit $n\le i <N$ is not valid anymore. 
As such, as $n$ decreases, the effective $Q_{max}$ decreases as well, thus potentially worsening the error correction performance.
Nevertheless, the rate of decrease is very slow: for example, with $N=128$ and $n=64$, only 64 patterns out of $2^{13}$ are invalid. 
As such, the BER loss will be noticeable only for very small $n$.

\subsection{Reduced switching activity}

The decoder architecture presented in \cite{LUTGRAND} limits the power consumption related to switching activity by gating the clock for the pipeline stages carrying unused information, for example the permutation $\pi$ after the valid codeword has been retrieved.
In addition to that, the proposed decoder can substantially lower the power consumption of stage 0 by deactivating the sorter as soon as $\mathbf{\hat{y}}^0_{vld}$ is asserted, i.e. as soon as ${\rm HD}(\mathbf{y})$ is identified as a valid codeword.
As channel conditions improve, the probability of receiving a valid codeword increases as well, being close to 50\% at SNR $=$ 6dB for $n=127$.

\section{Implementation}\label{sec:impl}

\ctable[
	caption={Proposed decoder implementation results versus \cite{LUTGRAND, ORBCHIP, ORB_gross}.},
	star,
	label={tab:BCH}, 	
    ]{l|rr|r|r|r}{
    }{
    	\toprule
	&  \multicolumn{2}{c|}{This work, $T_{fill}=4$} & \cite{LUTGRAND}.{\bf B} & \cite{ORBCHIP} & \cite{ORB_gross} \\
\midrule
		Result type & \multicolumn{2}{c|}{Synthesis} & Synthesis & Fabricated & Synthesis \\ 
		Technology [nm]  &  3 & 7 & 7 & 40 & 65\\
		Supply [V]  &  0.4 & 0.5 & 0.5 & 1.0 & 0.9 \\
		$n$ 	 	& Up to 128 	& Up to 128   &	Up to 128       & Up to 256 &  128  \\
		$R_{min}$   & 0.656 & 0.656			  &	0.656     & NA &  0.75  \\
		$r$         & 113/127	& 113/127		  &	113/127   & 240/256 & 105/128   \\
		$Q_{max}$	& $2^{13}$ 	& $2^{13}$		  &	$2^{13}$  & $\approx2^{23}$ &   $\approx2^{17}$ \\
		$Q_S$		& 512	& 512		  & 512       & - &  -  \\
		$T$			& 18 & 18		     & 18		  & - &  -  \\	
		$Q_{LUT}$	& 32 	& 32		  & 512       & - &  -  \\
	\midrule
 		$f$ [MHz]   &  845	&  616	& 616       &  90  & 454 \\
		Area [mm$^2$] &	1.98	&   3.14	& 3.70 & 0.4 & 1.82\\
	   \midrule
		A.C. $\mathcal{L}$  ~\multirow{2}{*}{[clock cycles] - [ns]}  & 11 - 13.02  & 11 - 17.86 	 & 25 - 40.58   & 5.5 - 61.3& 1.12 - 2.47\\
		W.C. $\mathcal{L}$  & 25 - 29.59  & 25 - 40.58  & 25 - 40.58   & NA & 42223 - 9300\\
    	\midrule
		A.C. $\mathcal{T}$  ~\multirow{2}{*}{[$\frac{\rm bits}{\rm cycle}$] - [Gb/s]}  &  113 - 95.49  &113 - 69.61  &113 - 69.61 & 47.8 - 4.3 & 93.6 - 42.5\\
		W.C. $\mathcal{T}$   & 113 - 95.49   & 113 - 69.61  &113 - 69.61 & NA & 0.025 - 0.011\\
    	\midrule
		Area eff. [Gbps/mm$^2$] & 48.23    &  22.17  & 18.81 & 10.75 & 23.3\\	
		Power [mW]  			& 101.41 & 158.56  &196.67  & 4.8 &104.3 \\
		Energy/bit [pJ/bit] & 1.16	&  2.28 &  2.83   & 1.14 & 2.45\\ 
		\bottomrule
}

The decoder architecture proposed in Section \ref{sec:newarch} has been synthesized in TSMC 3 nm FinFET technology typical corner, and the results are reported in Table \ref{tab:BCH} along with other state of the art ORBGRAND decoders.
The implementation parameters of the proposed decoder have been chosen to match those of architecture {\bf B} in \cite{LUTGRAND}. 
It supports any code length $n \le 128$ and any code rate $\ge 0.656$, and a maximum number of codebook queries $Q_{max}=2^{13}$, divided over a total of $T=18$ stages with $Q_S=512$. 
Following the modifications and analysis detailed in Section \ref{sec:newarch}, the newly introduced $T_{fill}$ is set to 4 and $Q_{LUT}$ is now limited to 32.
In 3nm technology, the achievable frequency $f$ is 845 MHz: the decoder guarantees a fixed throughput $\mathcal{T}$ of 95.49Gb/s, whereas latency $\mathcal{L}$ is 13.02ns on average (A.C), and 29.59ns in the worst case (W.C). 
It occupies 1.98 mm$^2$ and has a power consumption of 101.41 mW.
For a fair comparison with \cite{LUTGRAND}, the same architecture has been implemented also in TSMC 7nm technology. 
It can be seen that the proposed improvements account for 19.4\% gain in power and 15.2\% gain in area, while granting a 56\% reduction in average latency, at no expense in terms of error-correction performance (except for very small $n$), worst-case latency, or throughput.

The vast difference in technology nodes prevents any quantitative comparison with the other existing ORBGRAND decoders in literature \cite{ORBCHIP,ORB_gross}, since area and power scaling becomes more and more imprecise as the gap increases. 
Nevertheless, it can be seen that this work, similarly to \cite{LUTGRAND}, guarantees a high constant throughput and short worst case latency, while \cite{ORBCHIP,ORB_gross} focus on average performance and smaller footprint at the expense of worst-case metrics.

\section{Conclusion}\label{sec:conc}
In this work, a universal decoder architecture based on ORBGRAND with high constant throughput and short average and worst case latency has been proposed.
Its guaranteed worst case performance, along with the ability to exploit channel conditions for improved average latency and power consumption, make it a good candidate for demanding communication scenarios like 6G xURLLC.
The decoder output can be selected from any pipeline stage, with priority being given to valid and older codewords: a programmable threshold allows to strike the desired balance between error-correction performance and average latency and power consumption.
Together with strict clock gating and a more selective decoding algorithm programmability, the decoder yields 15.5\% area reduction, 19.4\% power reduction, and 56\% average latency reduction with respect to a baseline.
The decoder has been implemented in 3nm FinFET technology, and achieves a constant throughput of 95.49Gb/s, with a worst case and average latency of 29.59ns and and 13.02ns, respectively, while decoding a BCH code of length 127 and rate 113/127.

\bibliographystyle{IEEEbib}
\bibliography{IEEEabrv,GRAND_bib}

\end{document}

%% file: figures/TotalErrDistr.tikz
\begin{tikzpicture}
  \pgfplotsset{
    label style = {font=\fontsize{8pt}{7.2}\selectfont},
    tick label style = {font=\fontsize{6pt}{7.2}\selectfont}
  }

\begin{axis}[
	scale = 1,
    xlabel={SNR [dB]}, xlabel style={yshift=-0.2em},
    ylabel={\% of total decoding attempts}, ylabel style={yshift=-0.75em},
    ylabel near ticks,
    grid=both,
    ymajorgrids=true,
    xmajorgrids=true,
    grid style=dashed,
    mark options=solid,
    ybar interval,
    ymax = 100,
    ymin = 0,
    width=1\columnwidth, height=5.8cm,
    thick,
	xmin=1,
	xmax=9,
	xtick={1,2,3,4,5,6,7,8,9},
    xticklabels={3.0,3.5,4.0,4.5,5.0,5.5,6.0,6.5,7.0},
    legend style={
      anchor={center},
      cells={anchor=west},
      mark options=solid,
      column sep= 2mm,
      font=\fontsize{7pt}{7.2}\selectfont,
    },
    legend to name=TotErr,
    legend columns=8,
]
%

\addplot[
	color=CornflowerBlue,
	fill,
]
table {
1   59.5
2   38.4
3   19.3
4   6.9
5   1.7
6   0.25
7   0.021
8   1.0162736084673617e-05
9
};
\addlegendentry{$T_{fill}$=3}

\addplot[
	color=MidnightBlue,
	fill,
]
table {
1    49.7
2    27.2
3    10.0
4    2.1
5    0.26
6    0.024
7    1.9e-5
8    1.725747637020048e-06
9
};
\addlegendentry{$T_{fill}$=4}

\addplot[
	color=Blue,
	fill,
]
table {
1  42.3
2  20.1
3  5.8
4  1.1
5  0.19
6  0.024
7  1.9e-5
8    1.725747637020048e-06
9
};
\addlegendentry{$T_{fill}$=5}

\addplot[
	color=Black,
	fill,
]
table {
1 36.4
2 15.5
3 4.4
4 1.0
5 0.19
6 0.024
7 1.9e-5
8    1.725747637020048e-06
9
};
\addlegendentry{$T_{fill}$=6}

\addplot[
	color=BlueGreen,
	fill,
]
table {
1  31.5
2  12.8
3  4.1
4  1.0
5  0.19
6  0.024
7  1.9e-5
8    1.725747637020048e-06
9
};
\addlegendentry{$T_{fill}$=7}

\addplot[
	color=Green,
	fill,
]
table {
1   27.7
2   11.5
3   4.0
4   1.0
5   0.19
6   0.024
7   1.9e-5
8    1.725747637020048e-06
9
};
\addlegendentry{$T_{fill}$=8}

\addplot[
	color=PineGreen,
	fill,
]
table {
1   24.9
2   11.1
3   4.0
4   1.0
5   0.19
6   0.024
7   1.9e-5
8    1.725747637020048e-06
9
};
\addlegendentry{$T_{fill}$=9}

\addplot[
	color=YellowGreen,
	fill,
]
table {
1 23.3
2 11.0
3 4.0
4 1.0
5 0.19
6 0.024
7 1.9e-5
8    1.725747637020048e-06
9
};
\addlegendentry{$T_{fill}$=10}

\addplot[
	color=SeaGreen,
	fill,
]
table {
1  22.6
2  10.9 
3  4.0
4  1.0
5  0.19
6  0.024
7  1.9e-5
8    1.725747637020048e-06
9
};
\addlegendentry{$T_{fill}$=11}

\addplot[
	color=RoyalPurple,
	fill,
]
table {
1  22.3
2  10.9 
3  4.0
4  1.0
5  0.19
6  0.024
7  1.9e-5
8    1.725747637020048e-06
9
};
\addlegendentry{$T_{fill}$=12}

\addplot[
	color=Mulberry,
	fill,
]
table {
1 22.2
2 10.9  
3 4.0
4 1.0
5 0.19
6 0.024
7 1.9e-5
8    1.725747637020048e-06
9
};
\addlegendentry{$T_{fill}$=13}

\addplot[
	color=Sepia,
	fill,
]
table {
1 22.2
2 10.9  
3 4.0
4 1.0
5 0.19
6 0.024
7 1.9e-5
8    1.725747637020048e-06
9
};
\addlegendentry{$T_{fill}$=14}

\addplot[
	color=BrickRed,
	fill,
]
table {
1 22.2
2 10.9 
3 4.0
4 1.0
5 0.19
6 0.024
7 1.9e-5
8    1.725747637020048e-06
9
};
\addlegendentry{$T_{fill}$=15}

\addplot[
	color=BurntOrange,
	fill,
]
table {
1 22.2
2 10.9 
3 4.0
4 1.0
5 0.19
6 0.024
7 1.9e-5
8    1.725747637020048e-06
9
};
\addlegendentry{$T_{fill}$=16}

\addplot[
	color=Gray,
	fill,
]
table {
1  22.2
2  10.9 
3  4.0
4  1.0
5  0.19
6  0.024
7  1.9e-5	 
8    1.725747637020048e-06
9
};
\addlegendentry{$T_{fill}$=17}

%

\end{axis}
\end{tikzpicture}

%% file: figures/AddErrDistr.tikz
\begin{tikzpicture}
  \pgfplotsset{
    label style = {font=\fontsize{8pt}{7.2}\selectfont},
    tick label style = {font=\fontsize{6pt}{7.2}\selectfont}
  }

\begin{axis}[
	scale = 1,
    xlabel={SNR [dB]}, xlabel style={yshift=-0.2em},
    ylabel={\% of total decoding attempts}, ylabel style={yshift=-0.75em},
	ylabel near ticks, yticklabel pos=right,
    grid=both,
    ymajorgrids=true,
    xmajorgrids=true,
    grid style=dashed,
    mark options=solid,
    ybar interval,
    ymax = 100,
    ymin = 0,
    width=1\columnwidth, height=5.8cm,
    thick,
	xmin=1,
	xmax=9,
	xtick={1,2,3,4,5,6,7,8,9},
    xticklabels={3.0,3.5,4.0,4.5,5.0,5.5,6.0,6.5,7.0},
    legend style={
      anchor={center},
      cells={anchor=west},
      mark options=solid,
      column sep= 2mm,
      font=\fontsize{7pt}{7.2}\selectfont,
    },
    legend to name=AddErr,
    legend columns=4,
]
%

\addplot[
	color=CornflowerBlue,
	fill,
]
table {
1   37.3
2   27.5
3   15.3 
4   5.89
5   1.47
6   0.22
7   0.0195
8   8.62873818510024e-06
9
};
\addlegendentry{$T_{fill}$=3}

\addplot[
	color=MidnightBlue,
	fill,
]
table {
1 27.5
2 16.3
3 6.03
4 1.04
5 0.071
6 1.2e-05
7 0
8 
};
\addlegendentry{$T_{fill}$=4}

\addplot[
	color=Blue,
	fill,
]
table {
1 20.14
2 9.2
3 1.83
4 0.098
5 1.3790609514294734e-05
6 0
7 
};
\addlegendentry{$T_{fill}$=5}

\addplot[
	color=Black,
	fill,
]
table {
1 14.26
2 4.6
3 0.414
4 6.807351940095303e-05
5 0
6 
};
\addlegendentry{$T_{fill}$=6}

\addplot[
	color=BlueGreen,
	fill,
]
table {
1  9.35
2  1.8
3  0.071
4  5.236424569304079e-06
5  0
6  
}; 
\addlegendentry{$T_{fill}$=7}

\addplot[
	color=Green,
	fill,
]
table {
1 5.48
2 0.59
3 8.708272859216256e-05
4 0
5 
};
\addlegendentry{$T_{fill}$=8}

\addplot[
	color=PineGreen,
	fill,
]
table {
1  2.78
2  0.16
3  4.354136429608128e-05
4  0
5  
};
\addlegendentry{$T_{fill}$=9}

\addplot[
	color=YellowGreen,
	fill,
]
table {
1 1.15
2 0.036
3 1.4513788098693759e-05
4 0
5
};
\addlegendentry{$T_{fill}$=10}

\addplot[
	color=SeaGreen,
	fill,
]
table {
1  0.397
2  2.5994956978346202e-05
3  0
4  
};
\addlegendentry{$T_{fill}$=11}

\addplot[
	color=RoyalPurple,
	fill,
]
table {
1  0.116
2  0
3  
};
\addlegendentry{$T_{fill}$=12}

\addplot[
	color=Mulberry,
	fill,
]
table {
1   0.038
2   0
3   
};
\addlegendentry{$T_{fill}$=13}

\addplot[
	color=Sepia,
	fill,
]
table {
1  5.522573519259975e-05
2  0
3  
};
\addlegendentry{$T_{fill}$=14}

\addplot[
	color=BrickRed,
	fill,
]
table {
1 5.522573519259975e-05
2 0
3
};
\addlegendentry{$T_{fill}$=15}

\addplot[
	color=BurntOrange,
	fill,
]
table {
1   2.7612867596299876e-05
2   0
3
};
\addlegendentry{$T_{fill}$=16}

\addplot[
	color=Gray,
	fill,
]
table {
1  0
2
};
\addlegendentry{$T_{fill}$=17}

\end{axis}
\end{tikzpicture}

%% file: figures/BLER_Tfill.tikz
\begin{tikzpicture}[spy using outlines={circle, magnification=3.3, size=2cm, connect spies, transform shape}]
  \pgfplotsset{
    label style = {font=\fontsize{9pt}{7.2}\selectfont},
    tick label style = {font=\fontsize{7pt}{7.2}\selectfont}
  }

\begin{axis}[
	scale = 1,
    ymode=log,
    xlabel={SNR [\text{dB}]}, xlabel style={yshift=0.4em},
    ylabel={BLER}, ylabel style={yshift=-0.75em},
	ylabel near ticks, yticklabel pos=left,
    grid=both,
    ymajorgrids=true,
	yminorgrids=true,
	xmajorgrids=true,
    grid style=solid,
    mark options=solid,
    width=1\columnwidth,
    height=6.7cm,
	xtick={3,4,5,6,7},
    thick,
        xmin=3,
        xmax=7,
        ymin=1e-6,
		max space between ticks=20,
    mark size=3,
    legend style={
      anchor={center},
      cells={anchor=west},
      mark options=solid,
      column sep= 2mm,
      font=\fontsize{7pt}{7.2}\selectfont,
    },
    legend to name=BLER_tfill,
    legend columns=4,
]
%

\addplot[
    color=CornflowerBlue,
    mark=+,
    thick,
    mark size=3,
]
table {
3	0.786
3.5	0.498214
4	0.249
4.5	0.0847532
5	0.01959764
5.5	0.002828037
6	0.000269804
6.5	8.08629E-06
};
\addlegendentry{$T_{fill}$=3}

\addplot[
    color=MidnightBlue,
    mark=square,
    thick,
    mark size=3,
]
table {
3	0.688
3.5	0.386214
4	0.1563
4.5	0.0362532
5	0.00560764
5.5	0.000628037
6	7.48037E-05
6.5	0.000008
};
\addlegendentry{$T_{fill}$=4}

\addplot[
    color=BlueGreen,
    mark=x,
    thick,
    mark size=3,
]
table {
3 0.5065
3.5 0.241214
4 0.09671
4.5 0.025853252
5 0.00489764
5.5 0.000628037
6 7.48037E-05
6.5 0.000008

};
\addlegendentry{$T_{fill}$=7}


\addplot[
    color=Gray,
    mark=triangle,
    thick,
    mark size=3,
]
table {
3	0.413
3.5	0.223214
4	0.096
4.5	0.0258532
5	0.00489764
5.5	6.28E-04
6	7.48E-05
6.5	8.00E-06
};
\addlegendentry{$T_{fill}$=17}



\end{axis}
\end{tikzpicture}

%% file: figures/Complexity_Tfill.tikz
\begin{tikzpicture}[spy using outlines={circle, magnification=3.3, size=2cm, connect spies, transform shape}]
  \pgfplotsset{
    label style = {font=\fontsize{9pt}{7.2}\selectfont},
    tick label style = {font=\fontsize{7pt}{7.2}\selectfont}
  }

\begin{axis}[
	scale = 1,
    xlabel={SNR [\text{dB}]}, xlabel style={yshift=0.4em},
    ylabel={Average output stage $t$}, ylabel style={yshift=-0.75em},
	ylabel near ticks, yticklabel pos=right,
    grid=both,
    ymajorgrids=true,
	yminorgrids=true,
    xmajorgrids=true,
    grid style=solid,
    mark options=solid,
    width=1\columnwidth,
    height=6.7cm,
	xtick={3,4,5,6,7},
    thick,
        xmin=3,
        xmax=7,
        ymin=0,
    mark size=3,
    legend style={
      anchor={center},
      cells={anchor=west},
      mark options=solid,
      column sep= 2mm,
      font=\fontsize{7pt}{7.2}\selectfont,
    },
    legend to name=Comp_tfill,
    legend columns=8,
]

%

\addplot[
    color=CornflowerBlue,
    mark=+,
    thick,
    mark size=3,
]
table {
3	3
3.5 3
4	3
4.5 3
5	2.9999218532127523
5.5 2.999898
6	3
6.5	3
};
\addlegendentry{$T_{fill}$=3}

\addplot[
    color=MidnightBlue,
    mark=square,
    thick,
    mark size=3,
]
table {
3  3.999530581250863
3.5 3.996464685850945
4	3.9654571843251087
4.5 3.933057548306017
5   3.9742897069955165
5.5 3.995954
6	4
6.5	4
};
\addlegendentry{$T_{fill}$=4}

\addplot[
	color=Blue,
	thick,
]
table {
3   4.9967140687560403
3.5 4.9098494891990954
4	4.7266182873730043
4.5  4.8490234068178246
5   4.968715236171849
5.5 4.995937
6	5
6.5	5
};
\addlegendentry{$T_{fill}$=5}

\addplot[
	color=Black,
	thick,
]
table {
3   5.935220212619081
3.5 5.556759988562219
4	 5.468287373004354
4.5	5.8296172173639835
5	5.968454746881024
5.5 5.995937
6	6
6.5	6
};
\addlegendentry{$T_{fill}$=6}

\addplot[
    color=BlueGreen,
    mark=x,
    thick,
    mark size=3,
]
table {
3   6.677039900593677
3.5 5.9911097247134055
4	6.353062409288825
4.5	6.8280148714457765
5	 6.968454746881024
5.5 6.995937
6	7
6.5 7
};
\addlegendentry{$T_{fill}$=7}

\addplot[
	color=Green,
	thick,
]
table {
3	7.074306226701643
3.5 6.522680599963607
4	7.325181422351234
4.5	7.827747813792742
5	7.968454746881024
5.5 7.995937
6	8
6.5	8
};
\addlegendentry{$T_{fill}$=8}

\addplot[
	color=PineGreen,
	thick,
]
table {
3	7.176308159602375
3.5  7.269905638306168
4	8.319753265602322
4.5 8.827747813792742
5	8.968454746881024
5.5 8.995937
6	9
6.5	9
};
\addlegendentry{$T_{fill}$=9}

\addplot[
	color=YellowGreen,
	thick,
]
table {
3    7.2984950987160016
3.5 8.176661727624841
4	9.319013062409288
4.5 9.827747813792742
5	9.968454746881024
5.5 9.995937
6	10
6.5	10
};
\addlegendentry{$T_{fill}$=10}

\addplot[
	color=SeaGreen,
	thick,
]
table {
3   7.693414331078283
3.5 9.147937300163768
4	 10.31876632801161
4.5 10.827747813792742
5	10.968454746881024
5.5 10.995937
6	11
6.5	11
};
\addlegendentry{$T_{fill}$=11}

\addplot[
	color=RoyalPurple,
	thick,
]
table {
3	8.398149937871048
3.5  10.141750500402923
4	11.31876632801161
4.5 11.827747813792742
5	11.968454746881024
5.5 11.995937
6	12
6.5	12
};
\addlegendentry{$T_{fill}$=12}

\addplot[
	color=Mulberry,
	thick,
]
table {
3	9.272345713102306
3.5 11.140424757597026
4	12.31876632801161
4.5 12.827747813792742
5	12.968454746881024
5.5 12.995937
6	13
6.5	13
};
\addlegendentry{$T_{fill}$=13}

\addplot[
	color=Sepia,
	thick,
]
table {
3	 10.239016981913572
3.5 12.140424757597026
4	13.31876632801161
4.5	13.827747813792742
5	13.968454746881024
5.5 13.995937
6	14
6.5	14
};
\addlegendentry{$T_{fill}$=14}

\addplot[
	color=BrickRed,
	thick,
]
table {
3	11.231036863178241
3.5 13.140424757597026
4	 14.31876632801161
4.5 14.827747813792742
5	14.968454746881024
5.5 14.995937
6	15
6.5	15
};
\addlegendentry{$T_{fill}$=15}

\addplot[
	color=BurntOrange,
	thick,
]
table {
3    12.230567444429104
3.5 14.140424757597026
4	15.31876632801161
4.5 15.827747813792742
5	15.968454746881024
5.5 15.995937
6	16
6.5	16
};
\addlegendentry{$T_{fill}$=16}

\addplot[
    color=Gray,
    mark=triangle,
    thick,
    mark size=3,
]
table {
3	17
3.5 17
4	17
4.5 17
5	17
5.5 17
6	17
6.5	17
};
\addlegendentry{$T_{fill}$=17}
\end{axis}
\end{tikzpicture}